# PAKing up to the endothelium

Eva Maria Galan Moya[1,2,3], Armelle Le Guelte[1,2], Julie Gavard[1,2,*]

1. Institut Cochin, Universite Paris Descartes, CNRS (UMR 8104), Paris, France.
2. Inserm, U567, Paris, France.
3. Universidad de Castilla-La Mancha, Facultad de Medicina, Centro Regional de Investigaciones Biomedicas, Laboratorio de Oncologia Molecular.
* To whom correspondence should be addressed: Julie Gavard, 22 rue Mechain, Rm. 306, 75014 Paris, France. Ph: +33 1 4051 6424; Fax: +33 1 4051 6430; email: julie.gavard@inserm.fr

**Abstract:**
Angiogenesis recapitulates the growth of blood vessels that progressively expand and remodel into a highly organized and stereotyped vascular network. During adulthood, endothelial cells that formed the vascular wall retain their plasticity and can be engaged in neo-vascularization in response to physiological stimuli, such as hypoxia, wound healing and tissue repair, ovarian cycle and pregnancy. In addition, numerous human diseases and pathological conditions are characterized by an excessive, uncontrolled and aberrant angiogenesis. The signalling pathways involving the small Rho GTPase, Rac and its downstream effector the p21-activated serine/threonine kinase (PAK) had recently emerged as pleiotropic modulators in these processes. Indeed, Rac and PAK were found to modulate endothelial cell biology, such as sprouting, migration, polarity, proliferation, lumen formation, and maturation. Elucidating the Rac/PAK molecular circuitry will provide essential information for the development of new therapeutic agents designed to normalize the blood vasculature in human diseases.

## 1. General Overview

During embryonic development, blood vessels arise from endothelial precursors which share their origin with the hematopoietic lineage. These progenitors assemble into a primitive vascular labyrinth composed of small capillaries, upon a process called vasculogenesis. Then, along angiogenesis, vessels sprout off more side branches to colonize avascular areas in the embryo. Such vessels expand into a more mature vascular network of larger vessels ramifying into smaller branches [1]. During adulthood, endothelial cells remain mainly quiescent although angiogenesis can occur in response to physiological and pathological conditions [2]. Hence, elucidating the mechanisms underlying normal and abnormal vascular function is an exciting and important field of investigation as it may unveil new targets for the treatment of many human disease conditions, including cancer, ocular and inflammatory disorders, asthma, diabetes, acquired immunodeficiency syndrome, and bacterial infections. In this review, we will first delineate the cellular mechanisms involved in neo-vascularization and the molecular mechanisms of Rac/PAK signalling. Then, we will review how this signalling nexus impacts on developmental angiogenesis, endothelial plasticity and endothelial barrier function.

## 2. Molecular and cellular mechanisms of angiogenesis

### 2.1. Endothelial cell plasticity during angiogenesis

Two major developmental processes, namely vasculogenesis and angiogenesis, allow the embryonic vascular tree to establish. In vasculogenesis, endothelial cell precursors associate to compose a primitive vascular plexus. This step precedes angiogenesis, when new capillaries and vessels assembled from pre-existing ones. Postnatal neo-vascularization occurs primarily by angiogenesis, during wound healing, pregnancy, and ovarian cycle. This physiological process is often co-opted by tumour cells to build a new vascular network dedicated to supply oxygen and nutrients to the cancerous cells, thereby enabling them to proliferate, survive and metastasize [2]. Angiogenesis involves enlargement and remodelling of veins and arteries, together with sprouting, branching and complex networking of capillaries (Figure 1). Four steps can be identified along this process while all involved profound cellular remodelling: *i*) endothelial cell sprouting, *ii*) vessel outgrowth and guidance, *iii*) fusion and lumen formation, and, *iv*) maturation and perfusion [3]. Resident endothelial cells are the main cell target for the formation of new blood vessels, as they received pro-migratory, proliferative and survival signals [4]. Although still a matter of investigation, circulating progenitors from the bone marrow might also be recruited and engaged into the expansion of neo-vessels [5]. Finally, the surrounding differentiated extracellular matrix together with pericytes might provide molecular tracks for formation of a new vascular network [6]. Although more work is still required, endothelial cells constitute the main cellular target during neovascularisation processes.

### 2.2. Angiogenic factors: VEGF and beyond

Among the angiogenic factors involved developmentally in the angiogenesis process, the Vascular Endothelial Growth Factor (VEGF)/VEGF Receptor (VEGFR) and the Delta/Notch signalling axis are for now privileged molecular targets of anti-angiogenic therapies [7] (Table 1). Among the 5 members that compose the VEGF family in mammals (VEGF-A, -B, -C, -D and Placental Growth Factor, PlGF), VEGF-A emerges as the most important molecule identified so far to controls blood vessel morphogenesis. This ligand is required for differentiation of endothelial cell precursors, endothelial cell proliferation, sprouting and vascular remodelling, maturation and maintenance [8]. Keeping with this, the VEGF receptor VEGF-R2 (also known as KDR and Flk1) plays a key role to mediate VEGF-A intracellular signalling [9]. This tyrosine receptor kinase has been shown to activate multiple signalling pathways; such as MAPK (Mitogen Activated Protein Kinase), PKC (Protein Kinase C), PI3K (Phosphoinositide 3 kinase)/Akt, Src, and Rac, all of which might contribute to the endothelial plasticity regulation [10]. Interestingly, VEGF-R1 is rather involved in balancing VEGF-R2 functions in endothelial proliferation, and is therefore frequently proposed to oppose to VEGF-A pro-angiogenic signalling through VEGF-R2, although both can cooperate synergistically in pathological conditions [11-13]. Finally, VEGF-R3 biological function is mainly linked to lymphatic vessel maintenance [14]. Moreover, most of the alternate angiogenic pathways characterized so far have been shown to impact directly or indirectly on VEGF actions, by acting on VEGF expression and secretion, and on VEGF-R availability and activation. For all these reasons, VEGF-A and VEGF-R2 have been considered for more 15 years now as the ideal molecular targets in the development of anti-angiogenic strategies, notably in cancer treatment. To this regard, blockade of the VEGF pathway leads to blood vessel pruning and reduction of the endothelial sprouting [2]. Successful clinical trials had validated the medical use of anti-VEGF agents in the treatment of colorectal and breast cancer, as well as brain tumours and age-related macular degeneration.

On another hand, the Notch family of receptors initiate a signalling pathway, highly conserved throughout species with diverse roles in development. Their mode of activation implies a ligand-dependant cleavage of their extracellular domain and the subsequent release of the Notch intracellular domain (NICD) by a γ-secretase cleavage [7]. Once cleaved, this NICD translocates to the nucleus and activates target genes. Of interest, Notch function is linked to arterial differentiation [15, 16]. More recently, Notch and its ligand Delta4 had emerged as the master regulator of tip cell selection and migration during endothelial sprouting [17]. Interestingly, inactivation of the Delta4/Notch signalling axis in the tumour mass leads to the reduction of the tumour size, while paradoxically enhances vessel density, most likely by an aberrant increase in endothelial sprouting [18-20]. However, this vasculature is not functional, as no blood flow can be monitored [19]. Such mechanism uncouples tumour growth from neo-vascularization, and might therefore present an appealing therapeutic approach by normalization of the tumour vasculature, as oppose to strict anti-angiogenic strategies. Although genetic evidence had demonstrated a link between Notch and VEGF signalling [21], the exact molecular pathways involved downstream the Notch pathway in the tip cell selection and endothelial sprouting is still unclear. Finally, other growth factors and ligands of tyrosine kinase receptors, such as angiopoietins and Fibroblast Growth Factors (FGFs), as well as guidance cues and adhesion molecules were described to play a role during these processes (Table 1). However, for sake of space, they will not be developed here [4].

## 2.3. G protein coupled receptor (GPCR) signalling in angiogenesis

Many GPCR ligands directly trigger angiogenesis by acting on their cognate receptors expressed on the endothelial cell surface. Additionally, they can allow the recruitment of pro-angiogenic inflammatory cells and favour VEGF production and release in the microenvironment (Table 2). For example, the chemokine CXCL-8 (IL-8) through its receptor, CXCR-2 can elicit pro-angiogenic pathways through these three mechanisms: direct intracellular signalling, chemo-attraction of pro-inflammatory cells, and increase in VEGF local production [22, 23]. At the molecular level, intracellular pathways such as Akt/mTOR (mammalian target of rapamycin), MAPK, and Rho GTPases and the nuclear factors, HIF (hypoxia inducible factor), NFκB (nuclear factor κB) and STAT-3 (signal transducer and activator of transcription) could participate to angiogenesis and can be controlled through IL-8 stimulation in CXCR-2 expressing endothelial cells [23, 24, Koch, 1992 #33].

Sphingosine 1 phosphate (S1P) has also been characterized for its positive effect on proliferation, survival and migration of endothelial cells, as well as its ability to promote vessel stability. In addition, the knockout of its receptor S1P1-R results in embryonic lethality due to poor vascular development in mice [25]. Interestingly enough, a blocking anti-S1P antibody can efficiently block endothelial cell migration and resulting capillary formation, and, inhibit blood vessel formation induced by VEGF and FGF, as well as tumour-associated angiogenesis. This strategy also neutralized S1P-induced release of pro-angiogenic growth factors and cytokines [26]. In addition to IL-8 and S1P, we can mention thrombin, apelin, and SDF-1 among GPCR ligands with pro-angiogenic activities [27-29] (Table 2).

In conclusion, the formation of a hierarchical and highly organized network of blood vessels requires the orchestrated mechanisms of various signalling pathways. Among which, the Rac/PAK axis had emerged as a crucial mediator of cellular responses such as migration, polarity, adhesion, proliferation and takes a lead role in normal and abnormal angiogenesis.

### 3. Molecular mechanisms of the Rac/PAK signalling axis

#### 3.1. Molecular basis of Rac signalling

Rac is a Rho GTPase, which belongs to a family of over 20 molecules, involved in actin cytoskeleton dynamics. Similarly to most Rho GTPases, Rac switches between active GTP-bound and inactive GDP-bound forms. This cycling is regulated by three classes of proteins: guanine nucleotide exchange factors (GEFs), GTPase activating proteins (GAPs) and guanine nucleotide dissociation inhibitors (GDIs) (Table 3) [30, 31]. When bound to GTP, Rac interacts with and activates downstream effectors, thereby stimulating a plethora of processes, such as morphogenesis, neuronal development, cell migration, cell adhesion, phagocytosis, and differentiation.

Three isoforms, Rac1, Rac2 and Rac3, co-exist with different expression patterns. Whereas Rac1 is ubiquitous, Rac2 is restricted to the haematopoietic lineage and Rac3 is mostly present in brain tissues. Their expression profile might explain why despite high sequence homologies, Rac isoforms do not exhibit redundant functions in animal models. Historically, Rac has been involved in the regulation of the actin dynamics and the formation of lamellipodial structures and membrane ruffling [32]. These biological effects can be attributed to the ability of Rac to control the functions of actin nucleating proteins (such as WASP family verprolin-homologous protein and the formin family), actin capping proteins (gelsolin, cofilin), and membrane-associated actin binding proteins (spectrins) [33]. Altogether, this largely contributes to the Rac involvement in cell migration. Related to its function on the actin cytoskeleton, Rac participates in the formation, maturation and turn-over of cell-extracellular matrix and cell-cell adhesion in a wide range of cellular systems [32]. More specifically, Rac is known to control NADPH oxidase in neutrophil and macrophage and this process together with Rac-based phagocytosis can contribute to bacterial killing [34].

Another key role of Rac concerns the regulation of neuronal morphogenesis, including neurite outgrowth, neuron guidance and growth cone navigation [33]. Of note, numerous GEFs which exhibit a specific action on Rac activity over Rho are highly expressed in the central nervous system during development (Table 3). Given the mechanistic similarities between neuronal and vascular circuitries, it will be of high interest to dissect the individual roles of endothelial GEF in angiogenesis. To this regard, Vav2, Tiam, and Cool were demonstrated to play a role in the endothelial barrier function and permeability response to agents such as VEGF, S1P and lipopolysaccharides [35-37]. In addition, Vav orchestrates ephrin-based angiogenesis and VEGF-induced endothelial migration [38, 39].

#### 3.2. Structural and biochemical features of the PAK family

P21-activated kinases, PAKs, were the first identified binding partners of GTP-bound Rac and Cdc42, while have never been found to interact with GDP-bound form of these proteins [40]. PAKs are highly conserved serine/threonine kinases of six family members identified so far in eukaryotes (Figure 2). Based on biochemical and structural features, PAKs can be further classified in two groups: PAK-1, -2 and -3 form the group I and PAK-4, -5 and -6 constitute the group II [41].

Group I PAKs are characterized by an N-terminal region that includes a conserved PAK

binding domain (PBD, also named as CRIB domain for Cdc42 Rac interacting binding domain) which is involved in the interaction with Rac-GTP and Cdc42-GTP. This PBD overlaps with an auto-inhibitory domain (AID) [42]. This latter is engaged in PAK trans-dimerization with the C-terminal kinase domain, only in the absence of Rac signalling. Indeed, binding of Rac-GTP to the PBD triggers conformational changes and destabilization of the AID, therefore unleashing the kinase domain available for auto-phosphorylation and full activation [43]. Of note, this domain exhibits more than 90% of identity within group I kinases. In addition, PAKs bear canonical proline-rich Src Homology 3 (SH3) binding motifs within the N-terminal part that can recruit Nck and Grb2 sub-membrane adaptors, suggesting a role for membrane targeting in the modulation of PAK activity [41]. By contrast, the GEF PIX can directly bind to PAK through a non-typical proline-rich SH3 domain [44, 45]. Besides alternative splicing, diversities within group I PAK might come from a specific region that bridges between the two proline rich domains and that bear regulatory motif for phosphorylation and caspase cleavage [46]. Altogether, these biochemical properties contribute to the regulation of PAK activity and localization.

Although group II share the N-terminal PBD and the C-terminal kinase domain with group I, they notably lack other motifs found in PAK-1, -2, and -3 [41]. Importantly, they do not contain any AID, suggesting a different mode of activation. Indeed, despite the ability of group II PAK to interact with GTP-Cdc42 and to a lesser extent to GTP-Rac, their kinase activity is not significantly enhanced upon binding with activated Rho GTPases. Interestingly enough, all the substrates characterized so far for group I are also been found to be phosphorylated at least *in vitro* by group II PAKs [47]. In addition, crystal resolution of the three group II PAKs shows unique structural rearrangements along transition from a catalytically inactive open state to an active close state [41]. These features differ from those described for PAK-1, reinforcing the idea of divergent modes of regulation. Alternatively, group II activity could be regulated by their sub-cellular localization as it has been shown that Cdc42-GTP binding to group II PAK reroutes them to the Golgi apparatus, mitochondria and nucleus [48-50]. However, the exact role of Cdc42 in PAK compartmentalization remains unclear. These data indicate that the two groups of kinases are differently regulated and that instead of the auto-inhibition process, group II had favoured a regulatory process involving a yet to be defined mechanism.

### 3.3. Activation and Inactivation of PAK

How PAK-1 is activated has been extensively analyzed, as the crystal structures of both inactive and active kinase domain has been resolved [43, 51]. Based on these data, it has been concluded that PAK-1 exists as a dimer in a *trans* auto-inhibitory conformation in which the AID of one PAK molecule blocks the catalytic domain of the other. Binding of Rac-GTP or Cdc42-GTP induces a series of conformational changes, which starts with the disruption of the dimer and ends with the release of the kinase domain in a stable catalytically active conformation. In addition, phosphorylation of the Thr423 residue within the activation loop is critical to group I PAK activation [52]. Thr423 phosphorylation could be mediated either by auto-phosphorylation or by PDK1 (3-phosphoinositide-dependent protein kinase), suggesting membrane lipid signalling involvement. This residue, as well as auto-phosphorylation of several other sites, are exposed upon Rac/Cdc42 binding and contribute to kinase activation and stabilization (Figure 2).

Even though PAKs are considered downstream targets of active Rac/Cdc42, several Rho GTPase independent activation mechanisms have been reported. First, proteolytic cleavage of PAK removes its N-terminus domain involved in *trans* dimerization, while this proteolysis could enhance PAK activity *in vitro*. PAK-1 is for example cleaved and therefore activated by caspase 3 during apoptosis [53]. An additional important layer of regulation concerns PAK recruitment to the plasma membrane. To this regard several mediators have been reported: Nck interaction, C-terminal isoprenylation, PDK-1 phosphorylation, Akt dependent phosphorylation, upstream activation of Ras signalling and heterotrimeric G-protein β/γ [54-58]. Importantly, PDK1 and Akt activation of PAK can occur even in the presence of dominant negative form of Rac and Cdc42, demonstrating the Rho GTPase-independent activation of PAK. However, since the activating phosphorylations occur on sites that are masked in the inactive PAK dimers, it is still unclear how these motifs can be accessible for phosphorylation. Finally, PAK can directly bind to the GEF PIX/Cool family, which in turn recruits the G protein coupled receptor kinase-interacting target (GIT) and this scaffold ultimately leads to PAK activation [44]. This PAK-PIX-GIT complex accumulates at the plasma membrane, and again even in the absence of Rac or Cdc42 activation [59]. More specifically, this PAK-PIX-GIT complex localizes in focal adhesions in migrating cells [60]. In addition, GIT targets PAK to the centrosome in mitotic cells and induces Rho GTPase-independent activation of PAK. In all these cases, it is still not clear how PAK activation can be achieved in the absence of Rac/Cdc42 binding. Dimer breathing has been proposed, in which the kinase domain is temporarily released from the AID, and allows Rho GTPase-independent PAK activation in cooperation with the mechanisms mentioned above [51].

PAK inactivation represents also a physiological process important for cellular homeostasis. Indeed, PAK rapid kinetics of activation/inactivation suggests that its activity is tightly regulated. The initial activation of PAK by Rac/Cdc42 binding is rapidly blocked through GAP action. However, this process does not eliminate the initial burst of PAK activation which is subsequently exacerbated by auto-phosphorylation. Two serine/threonine phosphatases of the PP2C family, namely POPX1 and POPX2, can deactivate PAK mostly through Thr423 dephosphorylation [61]. Alternatively, active PAK was shown to be targeted for degradation through the proteasome via Chp and Cdc42 Rho GTPases, implying that these classes of proteins play contrasting roles as both activator and inhibitor of PAK signalling [62]. In addition, negative regulation of PAK can be achieved by interfering with its upstream activation, among the molecules involved, Caveolin, Nischarin (an α5β1-integrin-binding partner), CRIPAK (cysteine-rich inhibitor of PAK), hPIP (human PAK-interacting protein) and the tumour suppressor Merlin could prevent Rac/Cdc42 activation by direct binding to PAK [63]. It is noteworthy that these negative regulators can act via interactions with N- and C-terminus of PAK. However, no negative regulators of group II have been reported so far. Finally, even though some small molecule inhibitors have been reported to block PAK activity, the available ones are not specific enough as they can block upstream regulators of PAK such as Rac, Nck, MLK (myosin lineage kinase), and PDK1 [41]. Therefore, PAK RNA interference and gene disruption are for now the more useful approach to clearly decipher PAK involvement in any biological effects.

### 3.4. Biological effects of PAK signalling

The most well characterized function of PAK is the regulation of cytoskeletal organization, cell morphology and motility [64]. Indeed, it is notoriously known that activated PAK induced lamellipodia, filopodia and membrane ruffles, concomitantly with a loss of actin stress fibres, increased in focal adhesion turnover and motility (Figure 3). These biological effects mirror Rac function. In addition, PAK family proteins can also modulate polarity and traction forces, therefore contributing to oriented cell migration [65]. The molecular basis for these activities can be partially explained by a larger number of PAK substrates, with a role in cytoskeletal structure, such as actin binding proteins, signalling molecules, microtubule-regulating proteins. Of note, PAK is involved in lamellipodia extension and actin dynamics by direct phosphorylation of the p41-arc subunit of the actin assembling complex Arp2/3, the LIM kinase which acts on actin filament stabilization, as well as the regulatory myosin light chain (MLC), which is crucial for acto-myosin contractility and generation of traction forces during migration [65, 66]. Interestingly, some regulatory actions of PAK might be independent of its kinase activity. Indeed, the PAK-PIX-GIT complex is targeted to focal contacts and is implicated in their dynamics during polarized migration [59]. In addition, PAK is activated by integrin-mediated adhesion at the focal contacts, where it is recruited through PDGF and VEGF signalling [67]. This could be achieved either through Rac and Cdc42 activation or independently of them through physical interactions with Nck and PIX. This latter might involve sequential activation of paxillin linked to integrin, and then GIT that can bridge paxillin to PIX [59]. The precise function of the whole complex and the role of PAK is still matter of debate as it has been shown to induce disassembly and assembly of focal contacts, suggesting rather a role in adhesion turnover consistent with cell migration. Finally, PAK can also contribute to orchestrate polarized migration through its role on microtubule dynamics and its interaction to the polarity complex [68].

In addition to its established roles in cytoskeletal regulation, PAK-1 might also significantly participate to nuclear events and signalling (Figure 3). In interphase cells, a subset of cellular PAK accumulated in the nucleus, suggesting nuclear functions. Of note, PAK-1 directly associates with specific gene promoter and enhancer elements, exerting both positive and negative regulatory control on gene transcription [69]. Several transcription factors and transcriptional co-regulators are also PAK-1 substrates, including the forkhead family member FOXO1, the oestrogen and androgen receptors, the transcriptional repressors SHARP and C-terminal binding protein 1 (CTBP1), and Snail homologue 1 (SNAI1) [63]. To this regard, PAK role in hormone signalling have been recently unveiled. PAK-1 can directly phosphorylate the oestrogen receptor α and promote its trans-activation functions [70]. By this ligand-independent activation of the oestrogen pathway, PAK might thus contribute to the development of cancer cell resistance to anti-oestrogens, such as tamoxifen. In addition, PAK-6 was identified as an androgen receptor-interacting protein, which can inhibit androgen receptor-mediated transcriptional responses by a phosphorylation-dependent mechanism [71].

Furthermore, PAK leads a role in proliferation and survival (Figure 3). It is therefore not a surprise that gene amplification and overexpression of PAK proteins have been found in diverse human cancers [63]. For example, PAK overexpression can promote cell transformation through Cyclin D1 up-regulation and anchorage-independent cell growth [72]. Mammalian PAKs are also likely to be involved in the cell cycle progression. Indeed, PAK-1 phosphorylation and activity appear to be modulated upon cell cycle, as they increase before chromatin condensation [73]. More specifically, PAK-1 co-localizes with and phosphorylates histone H3 on condensing chromatin, suggesting an implication in nucleosome organization and compaction of DNA during mitosis [73]. In addition, PAK was found to localize at spindle poles, along the spindle apparatus itself and at centromeres in early mitosis. As mitosis progresses, PAK becomes associated with the spindle mid-body, and finally the contractile ring during cytokinesis, where small GTPases are known to have regulatory roles [73]. Of note, the phenotypes of mid-segregation, centrosome duplication, lack of spindle attachment and incomplete cytokinesis that occur with the deregulation of PAK share remarkable similarities with the Aurora kinases, a family of mitotic proteins that are activated by phosphorylation. Interestingly, PAK-1 can elicit the activation of Aurora A upon phosphorylation [60]. This results in the accumulation of Aurora A and influences in turn the formation of centrosomes. Therefore, either directly or indirectly, PAK-1 seems to have a regulatory role in chromosome condensation, duplication of the microtubule-organizing centre, spindle attachment and maturation and, possibly, the coordination of chromosome segregation and cytokinesis [60]. Finally, PAK might promote cell survival through phosphorylation-induced BAD (BCL2-antagonist of cell death) inactivation, leading to a pro-survival pathway [74]. Alternatively, inhibition of caspase 8 signalling-mediated apoptosis had also been reported to involve PAK [74].

The role of group I PAKs in MAPK signalling has been extensively studied, and it is well documented that PAKs activate the MAPK kinase, MEK1, by phosphorylation, as well as Raf [56]. Interestingly, inhibition of PAK function blocks the activation of MAPK by PDGF, but not by the epidermal growth factor EGF [75]. The role of the group II PAK in MAPK signalling is more controversial, as they have been reported to either activate or inhibit ERK pathway [48, 76]. The involvement of PAK in MAPK signalling had profound effects on abnormal cell proliferation, notably in cancers [63].

Studies on molecular signalling induced by angiogenic factors and loss-of-function in diverse animal models had led to the conclusion that the Rac/PAK axis can serve as important regulator of endothelial plasticity during developmental and adult angiogenesis, as well as for blood vessel quiescence. This will be reviewed in the next paragraph.

**4. Rac/PAK implications in endothelial biology**
**4.1. Role in developmental angiogenesis**

Development of the vascular tree is temporally and spatially regulated via two separate and sequential mechanisms: vasculogenesis and angiogenesis. Vasculogenesis is characterized by *de novo* formation of a primitive vessel plexus in developing organisms, although it may occur in the adult animal under special conditions. On the other hand, angiogenesis is a process of formation of new capillaries sprouting from the pre-existing vessels. Several studies have been carried out to manipulate Rac and PAK activity and level during development of the vascular network.

To this regard, Rac1 was recently proposed to exert an essential function in vascular development, as demonstrated by the engineering of conditional knockout Cre/Flox mice in the endothelial territories [77]. Indeed, such deletion results in embryonic lethality at mid-gestation and *Rac1*-deficient embryos exhibited a defective development of major vessels. Vascular development was completely absent in the yolk sacs from *Rac1* knockout mice. *In vitro* inactivation of *Rac1* hampers migration, tubulogenesis, adhesion and endothelial barrier function [77]. These vascular defects appear even more dramatic than those expected due to the exclusive role of Rac1 in cell migration. Altogether, Rac1 activity in endothelial cells is essential for vascular development, and conversely demonstrates that the deletion of *Rac1* in endothelial cells prevents developmental angiogenesis.

Rac1 involvement occurs early in the developmental process of vascular network organization. Indeed, Rac1-dependent migration and sprouting of endothelial cells is mediated by acto-myosin contractility at endothelial junctions and is abrogated by VE-cadherin-dependent cell-cell adhesion [78]. Conversely, blocking VE-cadherin function in organotypic angiogenesis assay and in Zebrafish embryos stimulates sprouting, suggesting that VE-cadherin plays a crucial role in the cessation of sprouting through its effect on Rac1 [78]. This implication of Rac1 in endothelial cell sprouting is consistent with its requirement for the generation of small branching vessels during vascular development [77]. Hence, unsuccessful suppression of endothelial cell sprouting may have a contribution to vascular remodelling defects observed in VE-cadherin null embryos [79].

The developmental roles of the PAK family members in angiogenesis are still not completely understood. Pioneer experiments using cultured endothelial cells have suggested a role for PAKs in angiogenesis [45]. More recently, a chicken chorioallantoic membrane (CAM) model has been developed to study the role of PAK-1 in *in vivo* angiogenesis [80]. This elegant system uses bFGF-soaked filters placed on the chorioallantoic membrane of developing chick embryos. Importantly, a peptide mimicking the effects of a dominant negative form of PAK-1 was sufficient to impair bFGF-elicited angiogenesis in this model [81]. On the other hand, PAK-2a mutation in Zebrafish results in embryonic brain haemorrhage with intact gross development of the vasculature and normal hemostatic function [82]. The authors could establish that PAK-2a, as well as PAK-2b, are required for the formation of a stable vasculature, at least in certain vessel territories [82]. Another member of the family, PAK-4, has been recently shown to have a developmental role in angiogenesis. Indeed, PAK-4 null embryos show abnormalities in both yolk sacs and placentas, as lack of vasculature throughout the extra-embryonic tissue and abnormally formed labyrinthine layer of the placenta [83].

Collectively, these data suggest that Rac1 and PAK could be potential therapeutic targets for numerous human diseases that involve pathological neovascularization.

### 4.2. Endothelial migration

Endothelial cell migration is essential for vasculogenesis and angiogenesis during development, as well as for wound repair of injuries along the vascular endothelium in adulthood. To carry out this process, cells must adhere and de-adhere to the substratum in a coordinated and polarized fashion. This process is orchestrated by the following steps: extension of filopodia and lamellipodia at the leading edge, formation of new adhesions in these regions, and detachment of adhesion contacts at the trailing edge of the cell [84]. Both involve profound remodelling of the actin cytoskeleton. It is known for more than ten years now that the formation of lamellipodia and filopodia is regulated through the effects of Rac and Cdc42 on the cytoskeleton, respectively [32]. Inactivation of Rac1 results *in vitro* in defective endothelial cell migration, adhesion to substratum, and organization of intercellular junctions, altogether suggesting a role for Rac1 in both plastic and quiescent states of the vasculature [77]. In addition, VEGF causes endothelial cell migration by inducing Rac1-dependent lamellipodia extension, and this effect is mediated by the GEF Vav2 [38]. Moreover, decrease expression of Vav2 by siRNA hampers both Rac activation and cell migration in response to VEGF stimulation in macrovascular endothelial cells [35, 38]. At the molecular level, VEGF can favour tyrosine phosphorylation of Vav2 and therefore activation of this GEF, while this phosphorylation can be prevented by inhibiting either VEGFR-2 activity or Src kinase activity and expression [35, 38]. Therefore, VEGF-induced Vav2 tyrosine phosphorylation and downstream activation of Rac1 depend on Src kinase activity and this VEGF/Src/Vav2/Rac signalling axis is involved in endothelial migration and wound closure.

Between the numerous effectors identified for Rac and Cdc42, PAK is the best characterized regarding to the regulation of the cytoskeleton dynamics, although its exact role is still under current investigation. Constitutively active mutants of PAK, as well as dominant negative forms of PAK have been engineered to elucidate the role of PAK in microvascular endothelial cell migration [45]. In such conditions, proper regulation of PAK is required for cell migration in microvascular endothelial cells, as either unregulated increase or decrease of PAK activity inhibits cell motility [45]. In addition, PAK can alter cell morphology and cytoskeletal organization. Although expression of a dominant form of PAK does not affect lamellipodia extension and membrane ruffling, such cells tend to spread more extensively than control cells [45]. On the other hand, active PAK-expressing cells show lamellipodia and ruffles similar to control cells, while PAK translocates less to this highly dynamic membrane region. However, dominant negative and active forms of PAK increase stress fibre formation and provoke larger focal adhesions, therefore provoking again similar effects [45]. Of note, the function of PAK might be subtler, as active PAK-expressing cells appear more dynamic than dominant negative-expressing ones [45]. This is reflected in the initiation of new contact points, the surface of cell retraction at the cell edge, as well as bending, shortening, and lengthening of actin cables [45], pointing for a role of the acto-myosin contractility machinery. Although PAK is not strictly required for formation of lamellipodia and filopodia through Rac and Cdc42, this kinase plays a key role in coordinating leading edge adhesion and trailing edge detachment in order to produce polarized cell movement. Accordingly, PAK function is necessary for endothelial cell migration.

Understanding the signalling pathways by which VEGF and other angiogenic factors stimulate Rac1 activation and cell migration and considering as well PAK as a target for pharmacological inhibitors of angiogenesis may be valuable to develop new therapeutic approaches to interfere with pathological angiogenesis.

### 4.3. Proliferation and survival

Apart from its role in migration, the Rac/PAK pathway had been linked to cell cycle progression, cell survival, proliferation and transformation by oncogenes [63, 85, 86]. For example, Ras but not Raf proto-oncogenes from the proliferative MAPK pathway, could activate PAK-1 in co-transfection assays in fibroblast cells. Testing multiple combinations of mutants within the canonical Ras/Rho GTPase/MAPK pathway revealed that proliferation by Ras, Rac and Rho requires PAK signalling to MAPK [86]. Moreover, a catalytically inactive PAK-1 mutant inhibits Ras transformation of fibroblast in two well-established assays for Ras transformation, namely focus formation and soft-agar assays [85]. Therefore, PAK signalling serves as a convergence point in transformation induced by Rho GTPases that are activated by mitogenic factors.

Transformation of endothelial cells by aberrant proliferation could lead to specific tumour formation and has been proposed to contribute to Kaposi sarcoma progression by the viral G protein coupled receptor (vGPCR) of the human herpes virus 8 [87]. Interestingly, vGPCR up-regulates secretion of critical cytokines through Rac1 activation, while inhibition of Rac1 reduced vGPCR tumorigenesis *in vivo* [88]. In addition, expression of dominant-negative mutants of PAK-1 inhibited vGPCR induced focus formation and growth in soft agar [89]. These results identify the Rac/PAK axis as a key mediator of vGPCR paracrine neoplasia, suggesting that this nexus may represent suitable therapeutic targets for the treatment of Kaposi sarcoma.

Under specific conditions, PAKs could block apoptosis, through BAD phosphorylation, Bim degradation, regulation of redox potential, and inactivation of the forkhead transcription repressor FKHR [90]. However, the specific role of PAK in endothelial cell survival has not been fully addressed so far.

### 4.4. Maturation and lumen formation

Although lumen formation of vessels is an essential step in the development of a functional vascular system, there is little understanding of how lumens originate in blood vessel. This is mainly due to the difficulty to establish appropriate *in vivo* models to allow the study of this event. However, a number of *in vitro* models have been used to clarify molecular requirements for endothelial cell lumen and tubule formation in a three-dimension extracellular matrix environment [91-93]. Endothelial cell lumen formation in three dimensional collagen matrices is regulated by the formation and coalescence of intracellular vacuoles, a process dependent on Cdc42 and Rac1 GTPases in response to integrin interaction [94, 95]. This represents a determinant mechanism of vascular development of Zebrafish, suggesting that these two Rho GTPases play a key role in vascular lumen formation *in vivo* [96].

Furthermore, PAK-2 and PAK-4 are also involved in endothelial cell lumen formation in three dimensional collagen matrices [94, 95, 97]. It has been shown that a multi-component kinase signalling pathway downstream of integrin-matrix interactions and Cdc42 activation controls endothelial cell lumen formation [97]. In conclusion, these models had helped to reveal novel regulators that control the signalling events mediating the crucial step of lumen formation in vascular morphogenesis. In this context, the role of two PAK family members has been emphasized, and implies both Rac/Cdc42 upstream regulation and complex signalling network interplay.

**4.5. Endothelial barrier function and quiescence**

In adulthood, fluids, cells, and nutrients exchange between the blood compartment and surrounding tissues can be increased or decreased, under physiologic conditions and depending on the vascular sites [98]. Thus, endothelial cells function as gatekeepers to control the infiltration of blood proteins and circulating cells to the surrounding microenvironment. This vascular permeability contributes to normal angiogenesis, blood pressure control, as well as immune responses. Many pathological conditions and human diseases, such as tumour-induced angiogenesis, inflammation, macular degeneration, allergy, and brain stroke, exhibit an abnormal vascular permeability increase. This function must therefore be tightly regulated to maintain the endothelial integrity.

Rac may have a dual role in the endothelial barrier function as it can contribute to both assembly and disassembly of VE-cadherin-based cell-cell junctions. Endothelial cells derived from VE-cadherin null embryos exhibit low levels of active Rac resulting from reduced junctional localization of the Rac exchange factor Tiam1 [99]. In addition, VE-cadherin signalling to Rac *via* Tiam1 stabilizes junctions and blocking Rac activity destabilizes junctions [78]. Conversely, Rac activation can impair endothelial barrier opening by permeability-inducing factors, such as thrombin [100]. Nevertheless, activation of Rac by VEGF in endothelial monolayer can destabilize junctions by promoting internalization of VE-cadherin [35]. This ultimately leads to increase in vascular permeability. Of note, blocking artificially the VEGF-induced Src-dependent Rac activation abrogates endothelial permeability increase, while similar mechanisms can be used by anti-permeability factors [35, 101, 102]. In addition, the platelet-activating factor and the IL-8 chemokine were recently shown to induce vascular permeability through Rac activation and endothelial cell junction disassembly [103, 104]. Finally, overexpression of dominant negative and active forms of Rac equally alters endothelial cell monolayer *in vitro*, provoking an increase in basal permeability [105]. This indicates that the level and the localization of Rac activation must be strongly regulated and suggest that there may be separate pools of activated Rac with distinct functions. The individual GEF involved might as well play a leading role in controlling Rac function. Hence the role of Rac in the endothelial barrier function is not fully elucidated and might imply a delicate control of the dynamics of the endothelial cell-cell junctions, and can eventually lead to either assembly or disassembly.

Altered vascular integrity and bleeding are frequently found in patients with vascular malformations. Of note, cerebral cavernous malformations (CCM) had helped to identify three classes of genes involved, and named CCM1-3. CCM2 interacts with both CCM1 and CCM3 and serves as a scaffold protein that binds to actin and the GTPase Rac [106]. In addition, leakiness of the vasculature, thin alignment of the endothelium and ultrastructure alteration of the endothelial junctions had been observed in brain tissue from CCM patients [107]. These results have been further demonstrated using mice and zebrafish models [108]. These clinical- and animal model-based data provided evidence for a link between Rac function and human diseases involving severe haemorrhages.

PAK could as well play a crucial role in controlling the endothelial barrier properties. In both bovine aortic and human umbilical vein endothelial cells, PAK has been found to be phosphorylated on Ser141 downstream of Rac activation, and this phosphorylated PAK fraction translocates to endothelial cell-cell junctions in response to serum, VEGF, bFGF, tumour necrosis factor alpha, histamine, and thrombin [109]. Blocking PAK activation or translocation prevents the augmentation of endothelial cell permeability in response to all these factors. In addition, permeability increase correlates with myosin phosphorylation, formation of actin stress fibres, and the appearance of paracellular pores [109]. These data suggest that PAK is a central regulator of endothelial permeability induced by multiple growth factors and cytokines *via* cell contractility. In addition, PAK might phosphorylate VE-cadherin in response to VEGF on a highly conserved serine residue which controls its stability at the plasma membrane; and therefore, provide a functional link between PAK activation and endothelial cell-cell junction remodelling in vascular permeability augmentation [35]. Finally, atherogenic flow profiles, oxidized low density lipoproteins, and pro-atherosclerotic cytokines all stimulate PAK phosphorylation and recruitment to cell-cell junctions, while inhibiting PAK *in vivo* reduces permeability in atherosclerosis-prone regions [110]. Similarly, blocking PAK function inhibits vascular leakage in a mouse model of acute lung injury caused by lipopolysaccharide treatment [36]. The molecular mechanisms involve the integrity of the multi-molecular complex PAK/PIX/GIT, such as a cell-permeant peptide, that blocks binding of PAK to PIX, inhibits inflammation-provoked vascular permeability in mouse lung [36]. Interestingly, PAK-2 mutations have been identified in Zebrafish embryo and lead to cerebral haemorrhage without alteration of brain vasculature [82]. This defect can be rescued by endothelial-targeted re-expression of wild-type PAK-2. These data correlate with another Zebrafish mutant in PIX exhibiting similar developmental defects and shown to act upstream of PAK-2 [111]. There is no question that PAK contributes uniquely to the onset of the vascular barrier during development and in the loss of barrier function in pathological conditions such as inflammation, atherosclerosis, allergy and tumour-induced angiogenesis. PAK may therefore be a suitable drug target for the treatment of human diseases where vascular permeability is exacerbated.

### 5. Concluding remarks

Tremendous efforts have been made towards a better understanding of the molecular mechanisms involved in normal and aberrant angiogenesis, since it bridges several exciting areas of research and opens new therapeutic opportunities for the treatment of many human diseases that involve pathological vessel leakiness, including acute and chronic inflammation, tissue damage following stroke and myocardial infarction, diabetic retinopathy, macular degeneration, and tumour-induced angiogenesis. In this context, the Rac/PAK signalling axis had emerged as a unique regulator of each aspect of the biology of the endothelium, as it can affect migration, polarity, survival, adhesion, and proliferation of the endothelial cells and contributes to the vasculature development. Although our present understanding of normal and pathological angiogenesis had helped to unveil Rac and PAK involvement, further studies are required to pinpoint the molecular mechanisms by which this pleiotropic nexus is acting. It is tempting to speculate that the development of specific inhibitors of Rac and individual PAK will further contribute to new treatment of human diseases and pathologies that are characterized by aberrant vascularization.

**Acknowledgements:**
We truly regret that we could not cite many seminal works owing to space limitations. We would like to thank N. Bidère (Inserm, Villejuif, France) for helpful discussions and comments. The authors are supported by funding from the Ligue Nationale contre le Cancer and by the IRG-People European program FP7. EMGM is supported by an individual fellowship from Consejeria de Sanidad de Castilla-La Mancha.

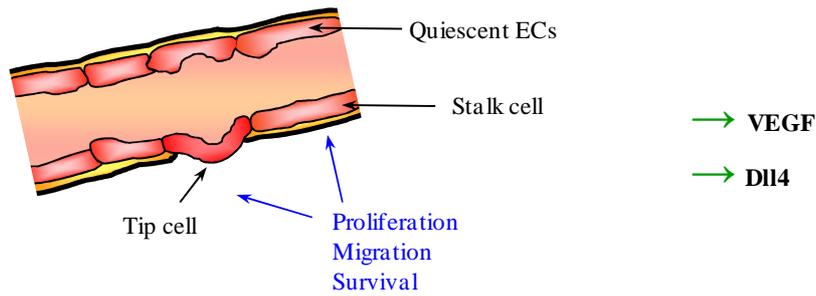
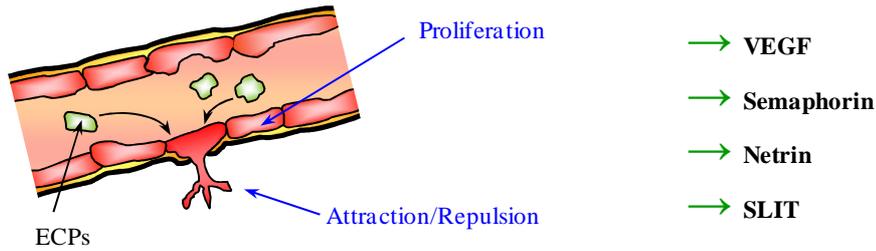
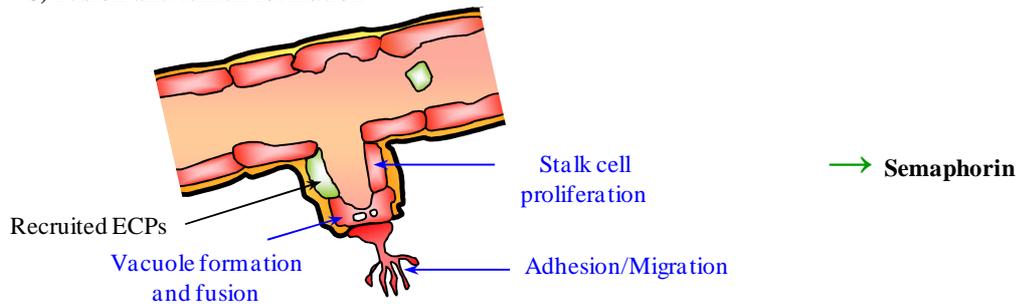
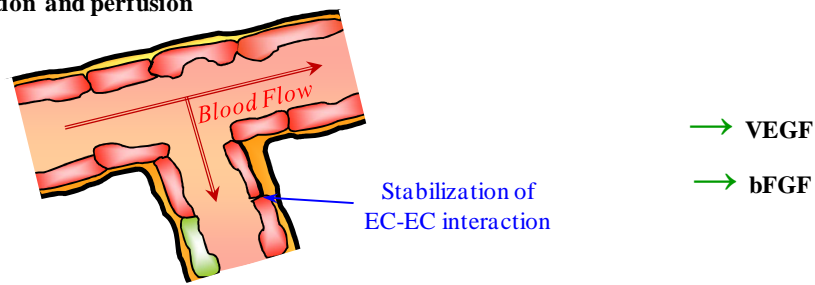

**Figure 1. Main steps of physiological angiogenesis.**
**A-** Formation of new blood vessels from resident endothelial cells (ECs, in red) is controlled upon pro-migratory, proliferative and survival signals, such as vascular endothelial growth factor (VEGF) and delta like-4 (Dll4). **B-** Vessel outgrowth is conducted by the orchestrated effects of angiogenic factors and guidance molecules among which VEGF, semaphorins, netrins and slit. Endothelial circulating progenitors (ECPs, in green) might contribute as well to neovascularisation. **C-** Fusion of EC vacuoles induces lumen formation in stalk ECs. Growth of new vessels is controlled by adhesive interactions acting on the migrating tip EC, in coordination with stalk EC proliferation. **D-** Once the new vessel is formed, EC-EC interactions are stabilized upon VEGF and basic fibroblast growth factor (bFGF) control. ECs, even still plastic, are entering in a quiescent state.

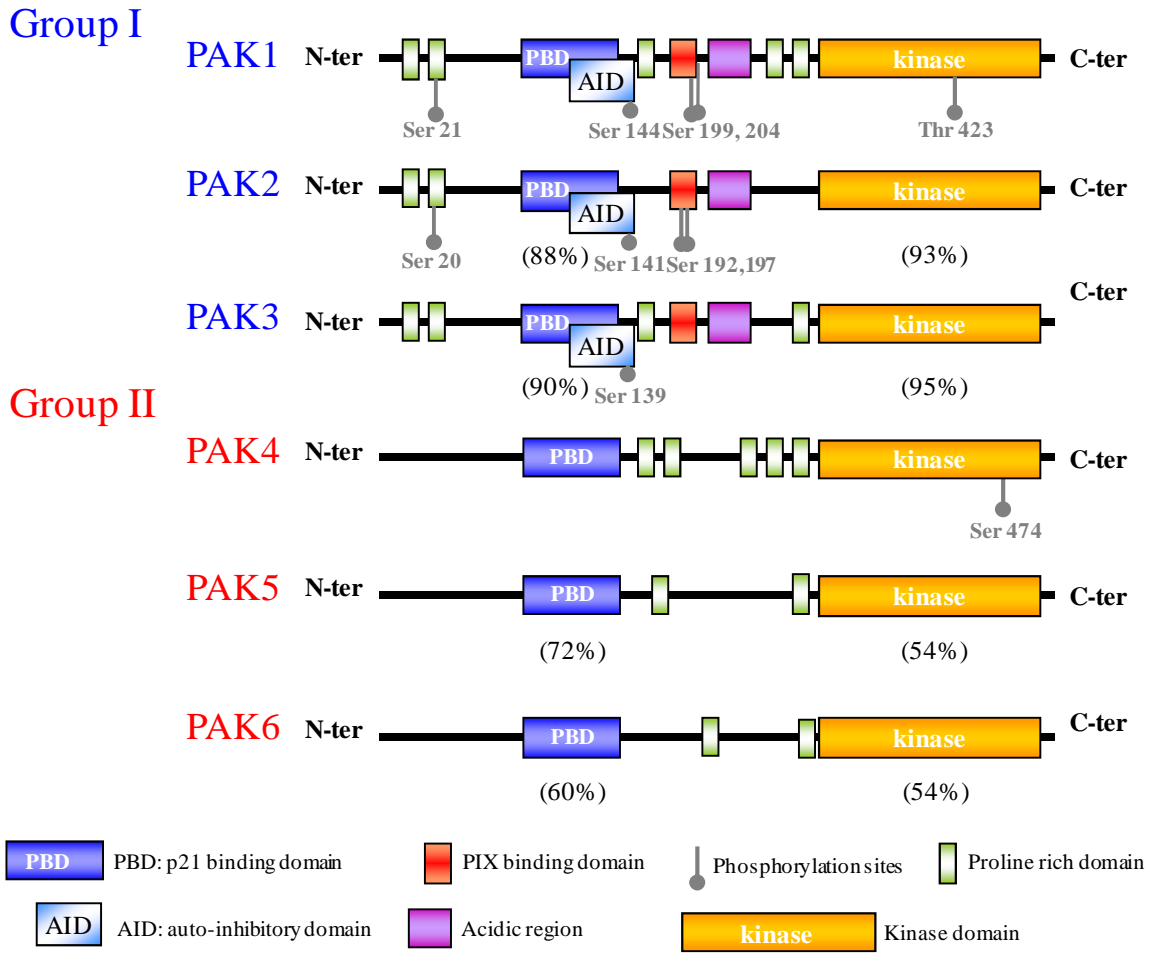

**Figure 2. Biochemical features of groups I and II p21-activated kinases.**
The six PAK family members are highly conserved serine/threonine kinases, as indicated by the percentage of identities below the two main domains, relative to PAK-1 for group I and to PAK-4 for group II. The group I is composed of a conserved PBD (PAK binding domain) that overlaps with an AID (auto-inhibitory domain) in the N-terminal region. In an inactive state, PAK is found in a close conformation, where PAK forms a dimer by trans-interaction between the kinase domain and the AID. GTP-bound Cdc42 and Rac1 interact with the PBD leading to rearrangement of the AID and therefore release of the catalytic domain. In addition, group I PAKs contain a proline-rich Src homology domain (SH3) binding motif which can recruit Nck and Grb2 adaptators and the GEF (guanine nucleotide exchange factor) PIX. Additionally, this group contains an acidic region that can be involved in protein-protein interaction. Diverse sites of phosphorylation have been described and modulate kinase activity and binding to the Nck adaptor protein. Group II includes a PBD that interacts with Rho GTPases but does not contain any AID, suggesting a divergent mode of activation. Unrelated conserved sequences are present. Within this group, the binding partners have not been clearly described yet, but might involve proline-rich sequences within the central core. Ser474 on PAK-4 is a phosphorylation site, analogous to Thr423 on PAK-1.

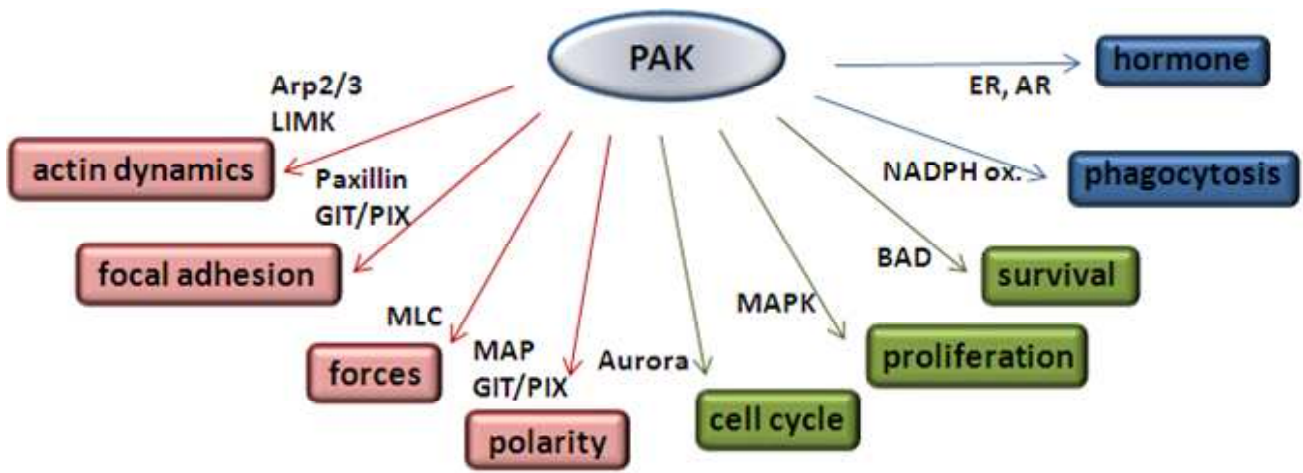

**Figure 3. Biological functions of PAK family kinases.**
A large number of PAK substrates has been described. PAK affected cell migration (red code) through effects on actin dynamics, focal adhesion turn-over, acto-myosin contractility and mechanical forces, as well as polarity. In addition, PAK has been shown to modulate cell proliferation and cell survival (green code). Finally, PAK can exert other atypical biological function (blue code) such as modulation of phagocytosis, and hormone signalling.

Arp2/3: actin assembling complex; LIMK: actin depolymerization inhibitor, LIM kinase; paxillin: integrin binding protein; MLC: myosin light chain; GIT/PIX: PAK scaffold proteins located in focal adhesions; MAP: microtubule associated proteins; Aurora kinase: mitotic regulating kinase; MAPK: mitogen associated protein kinase; BAD: Bcl-2 antagonist of cell death; NADPH ox.: nicotinamide adenine dinucleotide oxidase; ER, AR: oestrogen receptor, androgen receptor.

| Angiogenic Factor | Receptor | Cellular functions | Rho GTPase |
|---|---|---|---|
| VEGF-A | VEGF-R2 | Migration<br>Proliferation<br>Permeability<br>Differentiation<br>Survival | Rac1<br>Cdc42<br>RhoA |
| Dll-4 | Notch-1 | Tip selection<br>Arterial specification<br>Blockade of proliferation<br>Migration | ? |
| Angiopoietin-1 | Tie-2 | Survival<br>Migration<br>Vessel Stability | Rac<br>Rho |
| Semaphorins | Plexins<br>Neuropilins | Migration<br>Guidance | Rho |
| bFGF | FGF-R1, 2, 3, 4 | Migration<br>Survival<br>Proliferation<br>Differentiation<br>Vessel Stability | Rac |
| Slit-2 | Robo-4 | Vessel Stability | Rho<br>Rac<br>Cdc42 |
| Netrins | Unc5b | Guidance | Rac<br>Cdc42 |
| Ephrin | Ephrin receptors | Tip selection<br>Arterial specification<br>Proliferation<br>Adhesion | Rho |

**Table 1. Function of the main angiogenic factors in angiogenesis.**
This table lists the main angiogenic factors, their receptors and their respective roles in angiogenesis. Their ability to activate Rho GTPase (RhoA, Rac1 and Cdc 42) is also mentioned.

| Class | Protein Name | Target | Cellular function |
|---|---|---|---|
| GDI | Rho GDI-1 (α) | Rac RhoA Cdc42 | Binding to Rac and ERM complex Regulation NAPDH oxidase Exocytosis |
| | Rho GDI-2 (β, D4/Ly-GDI) | Rac RhoA Cdc42 | Inhibition of Rac without direct binding Inflammation in B/T lymphocytes |
| GAP | Chimaerins | Rac | Receptor for DAG Neuronal morphogenesis Migration and proliferation |
| | Oligophrenin | Rac RhoA Cdc42 | Neuronal morphogenesis Overexpressed in cancers |
| | 3BP1 | Rac | Ruffling |
| | Nadrin (Rich1,RhoGAP) | Rac Cdc42 | Ruffling Trafficking |
| | MgcRacGAP (RacGAP1) | Rac Cdc42 | Microtubule binding Mitotic spindle formation |
| | BCR, ABR | Rac Cdc42 | Regulation NAPDH oxidase Neuronal morphogenesis |
| | ARHGAP9 (RhoGAP9) | Rac Cdc42 | Overexpressed in hematopoietic cancers Adhesion |
| GEF | Sos 1, 2 | Rac | Ruffling, migration |
| | BCR, ABR | Rac RhoA Cdc42 | Neuronal morphogenesis Mutated in cancers |
| | RasGRF 1, 2 | Rac | Neuronal morphogenesis |
| | Ephexin (NGEF) | Rac RhoA Cdc42 | Neuronal morphogenesis |
| | Vav 1, 2, 3 | Rac Cdc42 RhoA | B/T lymphocyte function Cardio-vascular function Migration and proliferation |
| | α-, β-PIX (Cool 1, 2) | Rac Cdc42 | Migration Adhesion |
| | Tiam 1, 2 | Rac | Adhesion, migration |
| | PREX 1, 2 | Rac | Migration |
| | Asef | Rac | Ruffling, cell migration, cell adhesion |
| | Trio, Duo, Duet | Rac, RhoA | Neuronal morphogenesis |
| | DOCK | Rac Cdc42 | Neuronal morphogenesis Migration, adhesion |
| | SWAP | Rac | Adhesion, migration, trafficking |

**Table 3. Main regulators of Rac activity.**
Guanine nucleotide dissociation inhibitor (GDI) and GTPase activating protease (GAP) are both inhibitors of the Rac activity, while Guanine nucleotide exchange factor (GEF) positively contributes to Rac activity.